\documentstyle[twocolumn,aps,epsfig,floats]{revtex}

\begin{document}
	\wideabs{ %Do not forget to remove this line before submission.
	\title{Entropy production fluctuation theorem and the 
	nonequilibrium work relation for free energy differences}
\author{Gavin E. Crooks\footnotemark}
\address{Department of Chemistry, University of California at 
Berkeley, Berkeley, 
CA 94720}
	\date{\today}

	\maketitle
	
%---------  Abstract ---------------%
	\begin{abstract}
		 There are only a very few known relations in statistical 
		 dynamics that are valid for systems driven arbitrarily 
		 far-from-equilibrium.  One of these is the 
		 fluctuation theorem, which places conditions on the entropy 
		 production probability distribution of nonequilibrium 
		 systems.  Another recently discovered far-from-equilibrium 
		 expression relates nonequilibrium measurements of the work 
		 done on a system to equilibrium free energy differences.  In 
		 this paper, we derive a generalized version of the 
		  fluctuation theorem for stochastic, 
		 microscopically reversible dynamics.  Invoking this 
		 generalized theorem provides a succinct proof of the 
		 nonequilibrium work relation.
	\end{abstract}
	

	} %end wideabs. Remove before submission.
 \footnotetext[1]{Electronic address: gavinc@garnet.berkeley.edu}

	\section{Introduction}

	 Consider some finite classical system coupled to a constant 
	 temperature heat bath, and driven out of equilibrium by some 
	 time-dependent work process.  Most relations of nonequilibrium 
	 statistical dynamics that are applicable to systems of this sort 
	 are valid only in the linear, near-equilibrium regime.  One group 
	 of exceptions is the
	 entropy production fluctuation 
theorems\cite{Evans93,Evans94,GC95a,GC95b,Evans96,Gallavotti96,Cohen97,Gallavotti98,Kurchan98,Ruelle98,Lebowitz98,Maes98},
which are 
	 valid for systems perturbed arbitrarily far away from 
	 equilibrium.  Although the type of system, range of 
	 applicability, and exact interpretation differ, these 
	 theorems have the same general form,
	 
	\begin{equation}
			\frac{{\mathrm P}(+\sigma)}
				{{\mathrm P}(-\sigma)}
			\simeq e^{\tau \sigma}.
	\end{equation}
	
	\noindent Here ${\mathrm P}(+\sigma)$ is the probability of 
observing 
	an entropy production rate, $\sigma$, measured over a trajectory of 
	time~$\tau$.  Evans and Searles\cite{Evans94} gave a derivation for 
driven 
	thermostated deterministic systems that  are initially in 
equilibrium,
	Gallavotti and Cohen\cite{GC95a} rigorously derived their 
fluctuation theorem for 
	thermostated deterministic steady state ensembles, and 
	Kurchan\cite{Kurchan98}, Lebowitz and Spohn\cite{Lebowitz98}, and 
Maes\cite{Maes98} have 
	considered systems with stochastic dynamics.  The exact 
	interrelation between these results is currently under 
debate\cite{GC99,Evans99a}.
	 	 
	 In this paper we will derive the following, 
	 somewhat generalized, version of this theorem for stochastic 
	 microscopically reversible dynamics:
	
	\begin{equation}
			\frac{{\mathrm P_{F}}(+\omega)}
				{{\mathrm P_{R}}(-\omega)}
			= e^{+\omega}.
			\label{Eq:FT}
	\end{equation}
% 	\begin{equation}
% 			\frac{{\mathrm P_{F}}(+{\mathcal W})}
% 				{{\mathrm P_{R}}(-{\mathcal W})}
% 			= e^{+{\mathcal W}}.
% 			\label{Eq:FT}
% 	\end{equation}

	\noindent Here $\omega$ is the entropy production of the driven 
	system measured over some time interval, ${\mathrm 
	P_{F}}(\omega)$ is the probability distribution of this entropy 
	production, and ${\mathrm P_{R}}(\omega)$ is the probability 
	distribution of the entropy production when the system is driven 
	in a time-reversed manner.  This distinction is necessary because we 
	will consider systems driven by a time-dependent process, rather 
	than the steady perturbation considered elsewhere. The use of an 
entropy 
	production, rather than an entropy production rate, will prove 
convenient. 
	
	As a concrete example of a system for which the above theorem is 
	valid, consider a classical gas confined in a cylinder by a 
	movable piston.  The walls of this cylinder are diathermal so that 
	the gas is in thermal contact with the surroundings, which 
	therefore act as the constant temperature heat bath.  The gas is 
	initially in equilibrium with a fixed piston.  The piston is then 
	moved inwards at a uniform rate, compressing the gas to some new, 
	smaller volume.  In the corresponding time-reversed process, the 
	gas starts in equilibrium at the final volume of the forward 
	process, and is then expanded back to the original volume at the 
	same rate that it was compressed by the forward process.  The 
	microscopic dynamics of the system will differ for each repetition 
	of this process, as will the entropy production, the heat transfer, 
	and the work performed on the system.  The probability 
	distribution of the entropy production is measured over the 
	ensemble of repetitions.
	
	%*******Subsection: nonequilibrium work  relation********
		
	Another expression that is valid in the far-from-equilibrium 
	regime is the recently discovered relationship between the 
	difference in free energies of two equilibrium ensembles, $\Delta 
	F$, and the amount of work, $W$, expended in switching between 
	ensembles in a finite amount of 
time\cite{Jarzynski97a,Jarzynski97b,Jarzynski98a,Jarzynski98b,Crooks98a},
	
\begin{equation}
		\langle e^{-\beta W }\rangle = e^{-\beta \Delta F} .
		\label{Eq:Work}
	\end{equation}
	 \noindent Here $\beta = 1/k_{\mathrm B}T$, $k_{\mathrm B}$ is 
	 the Boltzmann constant, $T$ is the temperature of the heat bath 
	 that is coupled to the system, and $\langle \cdots\rangle$ 
	 indicates an average over many repetitions of the switching 
process.  For 
	 the confined gas considered above, the free energy depends on the 
	 position of the piston.  With Eq.~(\ref{Eq:Work}), we can 
	 calculate the free energy change when the system is compressed to 
	 a new volume by making many measurements of the work required to 
	 effect the change, starting each time from an equilibrated 
	 system, and taking the above average.  In the limit of 
	 instantaneous switching between ensembles, this relation is 
	 equivalent to the standard thermodynamic perturbation method used 
	 to calculate free energy differences with computer 
	 simulations\cite{Bennett76,Chandler87,Straatsma92,Frenkel96}.

	 Equations (\ref{Eq:FT}) and (\ref{Eq:Work}) are actually closely 
	 related.  We first note that whenever Eq.~(\ref{Eq:FT}) is valid, 
	 the following useful relation holds\cite[Eq.~(16)]{Gallavotti98}:
	
	\begin{eqnarray}
		\langle e^{-\omega}\rangle  &=&
	 		\int^{+\infty}_{-\infty}{\mathrm P_F}(+\omega) e^{-\omega} d\omega
	 			   =\int ^{+\infty}_{-\infty} {\mathrm P_R}(-\omega) d\omega 
	 			   \nonumber \\
	 			   &=& 1 .
		\label{Eq:One}
	\end{eqnarray}
	
	\noindent We shall show in Sec.~\ref{SecWP} that the generalized 
	fluctuation theorem, Eq.~(\ref{Eq:FT}), can be applied to systems 
	that start in equilibrium, and that the entropy production $\omega$ 
	for such systems is $-\beta\Delta F +\beta W$.  The nonequilibrium 
	work relation for the free energy change, Eq.~(\ref{Eq:Work}), can 
	thus be derived by substituting this definition of the entropy 
	production into Eq.~(\ref{Eq:One}), and noting that the free 
	energy difference is a state function, and can be moved outside 
	the average.

	In the following section we will derive the fluctuation theorem, 
	Eq.~(\ref{Eq:FT}).  Then in Sec.~\ref{SecWP} we will discuss two 
	distinct groups of driven systems for which the theorem is valid.  
	In the first group, systems start in equilibrium, and are then 
	actively perturbed away from equilibrium for a finite amount of 
	time.  In the second group, systems are driven into a time 
	symmetric nonequilibrium steady state.  We conclude by discussing 
	several approximations that are valid when the entropy production 
	is measured over long time periods.  The fluctuation theorem and 
	its approximations are illustrated with data from a simple 
	computer model.
		
	\section{The fluctuation theorem}
	As indicated in the Introduction, we consider finite, classical 
systems coupled to a 
	heat bath of constant temperature, $T=1/\beta$.  (All entropies 
	are measured in nats\cite{Shannon48}, so that Boltzmann's constant 
	is unity.) The state of the system is specified by $x$ and 
	$\lambda$, where $x$ represents all the dynamical, uncontrolled 
	degrees of freedom, and $\lambda$ is a controlled, time-dependent 
	parameter.  For the confined gas considered in the Introduction, 
	the heat bath is simply the walls of the cylinder, the state 
	vector $x$ specifies the positions and momenta of all the 
	particles, and $\lambda$ specifies the current position of the 
	piston.  In computer simulations, the controlled parameter could be 
	a microscopic degree of freedom.  For example, in a free energy 
	calculation $\lambda$ could specify the distance between two 
	particles or the chemical identity of an atom or 
	molecule\cite{Bennett76,Chandler87,Straatsma92,Frenkel96}.

	A particular path through phase space will be specified by the 
	pair of functions $\bigl(x(t),\lambda(t)\bigr)$.  It will prove 
	convenient to shift the time origin so that the paths under 
	consideration extend an equal time on either side of that origin, 
	$t\in\{-\tau,+\tau\}$ or $t\in\{-\infty,+\infty\}$.  Then the 
	corresponding time reversed path can be denoted as 
	$\bigl(\overline{x}(-t),\overline{\lambda}(-t)\bigr)$.  The overbar 
	indicates that quantities odd under a time reversal (such as 
	momenta or an external magnetic field) have also changed sign.

	%****** Subsection Dynamical assumptions *******

	 The dynamics of the system are required to be stochastic and
	 Markovian\cite{Grimmett92}.  The fluctuation theorem 
	 was originally derived for 
	 thermostated, reversible, deterministic 
systems\cite{Evans94,GC95a}. 
	  However, there are fewer technical 
	 difficulties if we simply assume stochastic 
dynamics\cite{Kurchan98}.
	 We will also require that the dynamics 
	 satisfy the following microscopically 
	 reversible condition:
	
	 \begin{equation}
	 	\frac{ {\mathcal P}[x(+t)|\lambda(+t)] }
	 		 { {\mathcal P}[\overline{x}(-t)|\overline{\lambda}(-t)]}
	 		 	= \exp\Big\{-\beta Q[x(+t),\lambda(+t)]\Big\}
	 		 	.
      		\label{Eq:MicroRev}
	\end{equation}
		
	\noindent Here ${\mathcal P}[x(+t)|\lambda(+t)]$ is the 
	probability, given $\lambda(t)$, of following the path $x(t)$ 
	through phase space, and ${\mathcal 
	P}[\overline{x}(-t),|\overline{\lambda}(-t)]$ is the probability 
	of the corresponding time-reversed path.  $Q$~is the heat, the 
	amount of energy transferred to the system from the bath.  The 
	heat is a functional of the path, and odd under a time reversal, 
	i.e., 
$Q[x(t),\lambda(t)]=-Q[\overline{x}(-t),\overline{\lambda}(-t)]$.  
	
	In current usage, the terms ``microscopically reversible'' and 
	``detailed balance'' are often used 
	interchangeably\cite{GrootMazur69}.  However, the original meaning 
	of microscopic reversibility\cite{Tolman24,Tolman38} is similar to 
	Eq.~(\ref{Eq:MicroRev}).  It relates the probability of a 
	particular path to its reverse.  This is distinct from the 
	principle of detailed balance\cite{GrootMazur69,Chandler87}, 
	 which refers to the probabilities of changing states without 
reference 
	to a particular path.  It is the condition that $ {\mathrm P}(A 
	\longrightarrow B)={\mathrm P}(B \longrightarrow A) \exp(-\beta 
	\Delta E)$, where $\Delta E$ is the difference in energy between 
	state $A$ and state $B$, and $P(A\longrightarrow B)$ is the 
	probability of moving from state $A$ to state $B$ during some 
	finite time interval. 
	 
	The stochastic dynamics that are typically used to model reversible 
	physical systems coupled to a heat bath, such as the Langevin 
	equation and Metropolis Monte Carlo, are microscopically 
	reversible in the sense of Eq.~(\ref{Eq:MicroRev}).  Generally if 
	the dynamics of a system obey detailed balanced locally in time 
	(i.e.  each time step is detailed balanced) then the system is 
	microscopically reversible even if the system is driven from 
	equilibrium by an external perturbation\cite[Eq~(9)]{Crooks98a}.
	
	%****** Entropy production *****

	%\newcommand{\ptau}{\!{\scriptscriptstyle +}\!\scriptstyle\tau}
	%\newcommand{\stau}{\mbox{\raisebox{-0.5pt}{-}}\!\scriptstyle\tau}

	A particular work process is defined by the phase-space 
	distribution at time $-\tau$, 
$\rho(x_{{\mbox{\raisebox{-0.5pt}{-}}\!\scriptstyle\tau}})$, and the 
value of the control 
	parameter as a function of time, $\lambda(t)$.  Each individual 
	realization of this process is characterized by the path that the 
	system follows through phase space, $x(t)$.  The entropy 
	production, $\omega$, must be a functional of this path.  Clearly
	there is a change in entropy due to the exchange of energy with 
	the bath.  If $Q$ is the amount of energy that flows out of the 
	bath and into the system, then the entropy of the bath must change 
by 
	$-\beta Q$.  There is also a change in entropy associated with the 
	change in the microscopic state of the system.  From an 
	information theoretic\cite{Shannon48} perspective, the entropy of a 
	microscopic state of a system, $s(x) =-\ln \rho(x)$, is the amount 
	of information required to describe that state given that the 
	state occurs with probability $\rho(x)$.  The entropy of this 
	(possibly nonequilibrium) \emph{ensemble} of systems is 
	$S=-\sum_{x} \rho(x) \ln \rho(x)$.  Thus, for a single realization 
	of a process that takes some initial probability distribution, 
	$\rho(x_{{\mbox{\raisebox{-0.5pt}{-}}\!\scriptstyle\tau}})$, and 
changes it to some different final 
	distribution, $\rho(x_{{\!{\scriptscriptstyle 
+}\!\scriptstyle\tau}})$, the entropy 
	production\cite{Footnote} is

		\begin{equation}
 		\omega = \ln 
\rho(x_{{\mbox{\raisebox{-0.5pt}{-}}\!\scriptstyle\tau}})
						-\ln \rho(x_{{\!{\scriptscriptstyle +}\!\scriptstyle\tau}})
						- \beta Q[x(t),\lambda(t)].
			\label{Eq:EntropyProduction}
		\end{equation}
			
	\noindent This is the change in the amount of information required 
	to describe the microscopic state of the system plus the change in 
	entropy of the bath.

	Recall that the fluctuation theorem, Eq.~(\ref{Eq:FT}), compares 
	the entropy production probability distribution of a process with 
	the entropy production distribution of the corresponding 
	time-reversed process.  For example, with the confined gas we 
compare 
	the entropy production when the gas is compressed to the entropy 
	production when the gas is expanded.  To allow this comparison of 
	forward and reverse processes we will require that the entropy 
	production is odd under a time reversal, i.e.,  $\omega_{\mathrm 
	F}=-\omega_{\mathrm R}$, for the process under consideration.  
	This condition is equivalent to requiring that the final 
	distribution of the forward process, $\rho_{_{\mathrm 
F}}\!(x_{{\!{\scriptscriptstyle +}\!\scriptstyle\tau}})$, 
	is the same (after a time reversal) 
	as the initial phase-space distribution of the reverse 
	process, $\rho_{_{\mathrm R}}\!(\overline{x}_{{\!{\scriptscriptstyle 
+}\!\scriptstyle\tau}})$,
	 and vice versa, i.e., 
	$\rho_{_{\mathrm F}}\!(x_{{\!{\scriptscriptstyle 
+}\!\scriptstyle\tau}})=\rho_{_{\mathrm 
	R}}\!(\overline{x}_{{\!{\scriptscriptstyle +}\!\scriptstyle\tau}})$
	and $\rho_{_{\mathrm 
R}}\!(x_{{\mbox{\raisebox{-0.5pt}{-}}\!\scriptstyle\tau}}) 
=\rho_{_{\mathrm 
F}}\!(\overline{x}_{{\mbox{\raisebox{-0.5pt}{-}}\!\scriptstyle\tau}})$.
In the next section, we will discuss two broad types of 
	work process that fulfill this condition.  Either the system 
	begins and ends in equilibrium or the system begins and ends in 
	the same time symmetric nonequilibrium steady state.

	This time-reversal symmetry of the entropy production allows the 
	comparison of the probability of a particular path, $x(t)$, starting 
	from some specific point in the initial distribution, with the 
	corresponding time-reversed path,
	\begin{equation}
		\frac{ \rho_{_{\mathrm 
F}}\!(x_{{\mbox{\raisebox{-0.5pt}{-}}\!\scriptstyle\tau}}) {\mathcal 
P}[x(+t)|\lambda(+t)]}
			{\rho_{_{\mathrm R}}\!(\overline{x}_{{\!{\scriptscriptstyle 
+}\!\scriptstyle\tau}}) {\mathcal 
			P}[\overline{x}(-t)|\overline{\lambda}(-t)]}
			= e^{+\omega_{\mathrm F}} .
		\label{anEq}
	\end{equation}
	This follows from the the conditions that the 
	system is microscopically reversible, Eq.~(\ref{Eq:MicroRev}) and 
	that the entropy production is odd under a time reversal.

	Now consider the probability, ${\mathrm P_{F}}(\omega)$, of 
	observing a particular value of this entropy production.  It can be 
	written as a $\delta$ function averaged over the ensemble of forward 
	paths,
	\begin{eqnarray*}	
				{\mathrm P}_{\mathrm F}(\omega) \!&=&
					\langle \delta(\omega {\raisebox{1.25pt}{\tiny $-$} } 
\omega_{\mathrm F})  \rangle_{_{\mathrm F}}\\ 
				&=&\!\!			 
\!\int\!\!\!\!\!\int\!\!\!\!\!\int_{x_{{\mbox{\raisebox{-0.5pt}{-}}\!\scriptstyle\tau}}}^{x_{{\!{\scriptscriptstyle 
+}\!\scriptstyle\tau}}}
			 \!\!\!\!\!\!\! \rho_{_{\mathrm 
F}}\!(x_{{\mbox{\raisebox{-0.5pt}{-}}\!\scriptstyle\tau}}) {\mathcal 
P}[x({\raisebox{1.25pt}{\tiny $+$} } 
t)|\lambda({\raisebox{1.25pt}{\tiny $+$} } t)] 
			\delta(\omega {\raisebox{1.25pt}{\tiny $-$} } \omega_{\mathrm F}) 
{\mathcal D}[x(t)] 
dx_{{\mbox{\raisebox{-0.5pt}{-}}\!\scriptstyle\tau}} 
dx_{{\!{\scriptscriptstyle +}\!\scriptstyle\tau}}
		.
	\end{eqnarray*}

	\noindent Here 
$\int\!\!\!\int\!\!\!\int_{x_{{\mbox{\raisebox{-0.5pt}{-}}\!\scriptstyle\tau}}}^{x_{{\!{\scriptscriptstyle 
+}\!\scriptstyle\tau}}} 
	\!\cdots {\mathcal{D}}[x(t)] 
dx_{{\mbox{\raisebox{-0.5pt}{-}}\!\scriptstyle\tau}} 
dx_{{\!{\scriptscriptstyle +}\!\scriptstyle\tau}}$ indicates a sum 
	or suitable normalized integral over all paths through phase 
	space, and all initial and final phase space points, over the 
appropriate 
	time interval. We can now use Eq.~(\ref{anEq}) to convert this 
	average over forward paths into an average over reverse paths.
	
		\begin{eqnarray*}
			{\mathrm P}_{\mathrm F}(\omega) \!&=&\!\!	
\!\int\!\!\!\!\!\int\!\!\!\!\!\int_{x_{{\mbox{\raisebox{-0.5pt}{-}}\!\scriptstyle\tau}}}^{x_{{\!{\scriptscriptstyle 
+}\!\scriptstyle\tau}}}
				\!\!\!\!\!\!\! \rho_{_{\mathrm 
R}}\!(\overline{x}_{{\!{\scriptscriptstyle +}\!\scriptstyle\tau}}) 
{\mathcal P}[\overline{x}({\raisebox{1.25pt}{\tiny $-$} } 
t)|\overline{\lambda}({\raisebox{1.25pt}{\tiny $-$} } t)] 
			 	 \times \nonumber\\*[-8pt]
			 	&&\qquad\qquad\qquad\qquad\delta(\omega {\raisebox{1.25pt}{\tiny 
$-$} } \omega_{\mathrm F}) e^{+\omega_{F}} 
				{\mathcal D}[x(t)] 
dx_{{\mbox{\raisebox{-0.5pt}{-}}\!\scriptstyle\tau}} 
dx_{{\!{\scriptscriptstyle +}\!\scriptstyle\tau}}\nonumber\\
			&=& e^{+\omega}\!\!	
\!\int\!\!\!\!\!\int\!\!\!\!\!\int_{x_{{\mbox{\raisebox{-0.5pt}{-}}\!\scriptstyle\tau}}}^{x_{{\!{\scriptscriptstyle 
+}\!\scriptstyle\tau}}}
				\!\!\!\!\!\!\! \rho_{_{\mathrm 
R}}\!(\overline{x}_{{\!{\scriptscriptstyle +}\!\scriptstyle\tau}}) 
{\mathcal P}[\overline{x}({\raisebox{1.25pt}{\tiny $-$} } 
t)|\overline{\lambda}({\raisebox{1.25pt}{\tiny $-$} } t)] 
			 	 \times \nonumber\\*[-8pt]
			 	&&\qquad\qquad\qquad\qquad\qquad\delta(\omega 
{\raisebox{1.25pt}{\tiny $+$} } \omega_{\mathrm R}) 
				{\mathcal D}[x(t)] 
dx_{{\mbox{\raisebox{-0.5pt}{-}}\!\scriptstyle\tau}} 
dx_{{\!{\scriptscriptstyle +}\!\scriptstyle\tau}}\nonumber\\
			 	&=& e^{+\omega} \,\langle \delta(\omega {\raisebox{1.25pt}{\tiny 
$+$} } \omega_{\mathrm R})  
			 	\rangle_{_{\mathrm R}}\nonumber\\ 
			 	&=& e^{+\omega} \,{\mathrm P_R}(-\omega).
		\end{eqnarray*}
	\noindent The $\delta$ function allows the $e^{+\omega}$ term to be 
moved 
	outside the integral in the second line above. The remaining average 
	is over reverse paths as the system is driven in reverse. The final 
result is the entropy 
	production fluctuation theorem, Eq.~(\ref{Eq:FT}).

	%******** Alternative Ensembles and Multiple Baths. ********
	The theorem readily generalizes to other ensembles.  As an example, 
	consider an isothermal-isobaric system.  In addition to the heat 
	bath, the system is coupled to a volume bath, characterized by 
	$\beta p$, where $p$ is the pressure.  Then the microscopically 
	reversible condition, Eq.~(\ref{Eq:MicroRev}), becomes

		\begin{eqnarray}
			\frac{{\mathcal P}[x(+t)|\lambda(+t)]}
					{{\mathcal P}[\overline{x}(-t)|\overline{\lambda}(-t)]}
			 = \exp\Big\{-\beta Q[&&x(t),\lambda(t)] \nonumber\\
			 &&-\beta p\Delta 
			 V[x(t),\lambda(t)]\Big\}.
			 \nonumber
		\end{eqnarray}

	\noindent Both baths are considered to be large, equilibrium, 
	thermodynamic systems.  Therefore, the change in entropy of the 
	heat bath is $-\beta Q$ and the change in entropy of the volume 
	bath is $-\beta p\Delta V$, where $\Delta V$ is the change in 
	volume of the system. The entropy production should then be defined 
as
		\begin{equation}
	 		\omega = \ln 
\rho(x_{{\mbox{\raisebox{-0.5pt}{-}}\!\scriptstyle\tau}})
						-\ln \rho(x_{{\!{\scriptscriptstyle +}\!\scriptstyle\tau}})
						-\beta Q -\beta p\Delta V .
				\label{Eq:EntropyProduction2}
		\end{equation}

	\noindent The fluctuation theorem, Eq.~(\ref{Eq:FT}), follows as 
	before.  It is possible to extend the fluctuation theorem to any 
	standard set of baths, so long as the definitions of microscopic 
	reversibility and the entropy production are consistent.  In the 
	rest of this paper we shall only explicitly deal with systems 
	coupled to a single heat bath, but the results generalize 
	directly.
	 	 
	\section{Two groups of applicable systems}\label{SecWP}				
	%\subsection{Systems that start in equilibrium.}

	\begin{figure}[t]
		\centering
		\epsfig{file=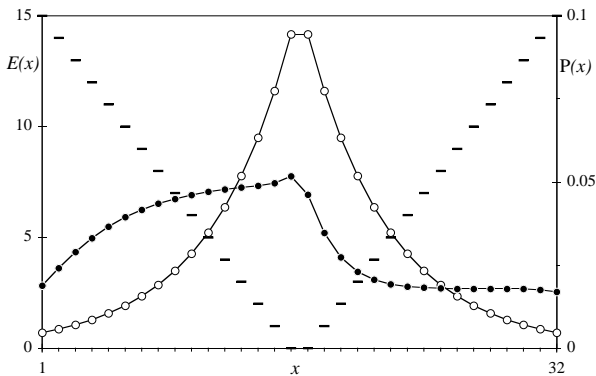, width=8cm} 
		\caption{A very simple 
		Metropolis Monte Carlo simulation is used to illustrate the 
		fluctuation theorem and some of its approximations.  The 
		master equation for this system can be solved, providing exact 
		numerical results to compare with the theory.  A single 
		particle occupies a finite number of positions in a 
		one-dimensional box with periodic boundaries, and is coupled 
		to a heat bath of temperature $T=5$.  The energy surface, 
		$E(x)$, is indicated by \protect\rule[0.5mm]{1.5mm}{0.4mm} in 
		the figure.  At each discrete time step the particle attempts 
		to move left, right, or stay put with equal probability.  The 
		move is accepted with the standard Metropolis acceptance 
		probability\protect\cite{Metropolis53}.  Every eight time steps, 
		the energy surface moves right one position.  Thus the system 
		is driven away from equilibrium, and eventually settles into a 
		time symmetric nonequilibrium steady-state.  The equilibrium 
		($\circ$) and steady state ($\bullet$) probability 
		distributions are shown in the figure above.  The steady state 
		distribution is shown in the reference frame of the surface.}
		\label{Fig:Profile}
	\end{figure}

	\begin{figure}[t]
		\centering
			\epsfig{file=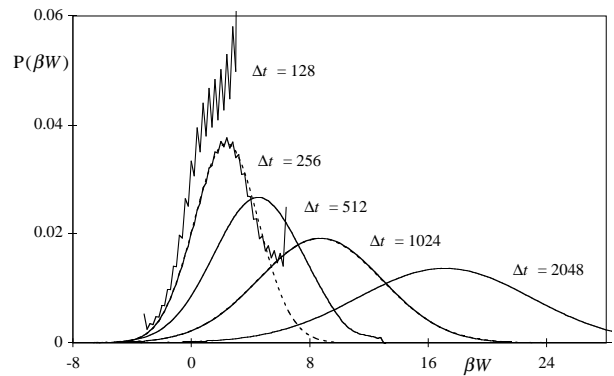, width=8cm}
			 \caption{Work 
			probability distribution (---) for the system of 
			Fig.~\ref{Fig:Profile}, starting from equilibrium.  The 
			work $W$ was measured over 16, 32, 64, 128, and 256 cycles 
			($\Delta t =$ 128, 256, 512, 1024, and 2048).  For each of 
			these distributions the work fluctuation theorem, 
			Eq.~(\ref{Eq:WorkFT}), is exact.  The dashed lines (- - -) 
			are Gaussians fitted to the means of the distributions for 
			$\Delta t =$ 256 and 1024.  (See Sec.~\ref{Sec:Approx}). }
		 	\label{Fig:Work}
	\end{figure}

	In this section we will discuss two groups of systems for which the 
	entropy fluctuation theorem, Eq.~(\ref{Eq:FT}), is valid.  These 
	systems must satisfy the condition that the entropy production, 
	Eq.~(\ref{Eq:EntropyProduction}), is odd under a time reversal, and 
	therefore that  $\rho_{_{\mathrm F}}\!(x_{{\!{\scriptscriptstyle 
+}\!\scriptstyle\tau}})=\rho_{_{\mathrm 
	R}}\!(\overline{x}_{{\!{\scriptscriptstyle +}\!\scriptstyle\tau}})$. 

	First consider a system that is in equilibrium from time 
	$t=-\infty$ to $t=-\tau$.  It is then driven from equilibrium by a 
	change in the controlled parameter, $\lambda$.  The system is 
	actively perturbed up to a time $t=+\tau$, and is then allowed to 
	relax, so that it once again reaches equilibrium at $t=+\infty$.  
	For the forward process the system starts in the equilibrium 
	ensemble specified by $\lambda(-\infty)$, and ends in the ensemble 
	specified by $\lambda(+\infty)$.  In the reverse process, the 
	initial and final ensembles are exchanged, and the entropy 
	production is odd under this time reversal.  The gas confined in 
	the diathermal cylinder satisfies these conditions if the piston 
	moves only for a finite amount of time.
				
	At first it may appear disadvantageous that the entropy 
	production has been defined between equilibrium ensembles 
	separated by an infinite amount of time.  However, for these 
	systems the entropy production has a simple and direct 
	interpretation.  The probability distributions of the initial and 
	final ensembles are known from equilibrium statistical mechanics:
		
	\begin{eqnarray}
		{\rho}_{\mathrm eq}(x|\lambda) & = & 
		\frac{e^{-\beta E(x,\lambda)}}
				{\displaystyle \sum_{x} e^{-\beta E(x,\lambda)}} \nonumber \\
		 & = & \exp\big\{\beta F(\beta,\lambda) -\beta E(x,\lambda)\big\} .
		\label{Eq:canonical}
	\end{eqnarray}

	\noindent The sum is over all states of the system, and 
	$F(\beta,\lambda) =-\beta^{-1}\ln \sum_{x}\exp\{-\beta 
	E(x,\lambda)\}$ is the Helmholtz free energy of the system.  If 
	these probabilities are substituted into the definition of the 
	entropy production, Eq.~(\ref{Eq:EntropyProduction}), then we find 
	that
		 
	\begin{equation}
		\omega_{\mathrm F} = -\beta \Delta F +\beta W .
		\label{Eq:OmegaWork}
	\end{equation}

	\noindent Here $W$ is the work and we have used the first law of 
	thermodynamics, $\Delta E = W + Q$.
		
	It is therefore possible to express the fluctuation theorem in 
	terms of the amount of work performed on a system that starts in 
	equilibrium\cite{Jarzynski97p},
			
	\begin{equation}
		\frac{{\mathrm P}_{\mathrm F}(+\beta W)}{{\mathrm P}_{\mathrm 
		R}(-\beta W)} 
		=e^{-\Delta F} e^{+\beta W} .\label{Eq:WorkFT}
	\end{equation}
		
	\noindent The work in this expression is measured over the finite 
	time that the system is actively perturbed.  We have now 
	established the fluctuation theorem for systems that start in 
	equilibrium, and have shown that the entropy production is simply 
	related to the work and free energy change.  Therefore, we have 
	established the nonequilibrium work relation, Eq.~(\ref{Eq:Work}), 
	discussed in the Introduction.

	The validity of this expression can be illustrated with the very 
	simple computer model described in Fig.~\ref{Fig:Profile}.  
	Although not of immediate physical relevance, this model has the 
	advantage that the entropy production distributions of this driven 
	nonequilibrium system can be calculated exactly, apart from 
	numerical roundoff error.  The resulting work distributions are 
shown in 
	Fig.~\ref{Fig:Work}.  Because the process is time symmetric, 
	$\Delta F=0$ and ${\mathrm P}(+\beta W) ={\mathrm P}(-\beta W) 
	\exp \beta W$.  This expression is exact for each of the 
	distributions shown, even for short times when the distributions 
	display very erratic behavior.

	 \begin{figure}[t]
		 	\centering
		 	\epsfig{file=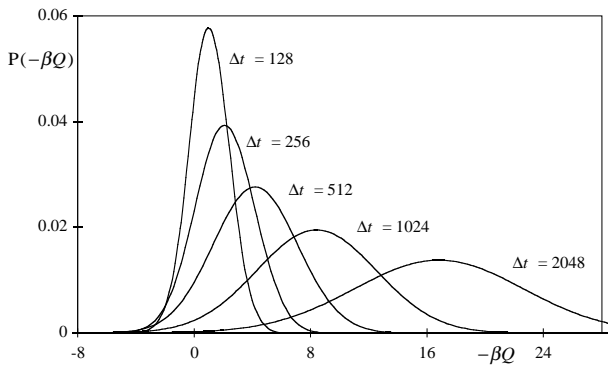, width=8cm} 
			\caption{ Heat probability distribution (---) 
			for the nonequilibrium steady state.  The model described in 
			Fig.~\ref{Fig:Profile} was relaxed to the steady state, 
			and the heat $Q$ was then measured over 16, 32, 64, 128 
			and 256 cycles ($\Delta t =$ 128, 256, 512, 1024, and 
			2048).  Note that for long times the system forgets its 
			initial state and the heat distribution is almost 
			indistinguishable from the work distribution of the system 
			that starts in equilibrium.}
		 	\label{Fig:Heat}
		\end{figure}

	\begin{figure}[t]
	 	\center 	
	 	\epsfig{file=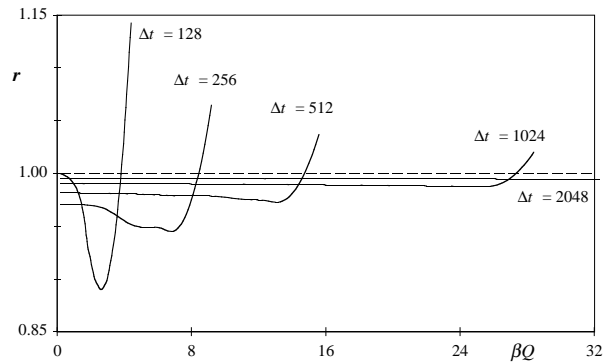, width=8cm} 
		\caption{Deviations from the heat fluctuation theorem, 
		Eq.~(\ref{Eq:HeatFT}), for the distributions of 
		Fig.~\ref{Fig:Heat}.  If the heat fluctuation theorem
		were exact, then the ratio 
		$r =\beta Q / \ln [ {\mathrm P}(+\beta Q)/{\mathrm P}(-\beta Q)]$ 
		 would equal 1 for all $|\beta Q|$.  For short 
		times ($\Delta t \leq 256$) the fluctuation theorem is 
		wholly inaccurate.  For times significantly longer than 
		the relaxation time of the system ($\Delta t \approx 100$), 
		the fluctuation theorem is accurate except for very large 
		values of $|\beta Q|$.}
	 	\label{Fig:Ratio}
	 \end{figure}
	
		\begin{figure}[t]
		 	\centering
			\epsfig{file=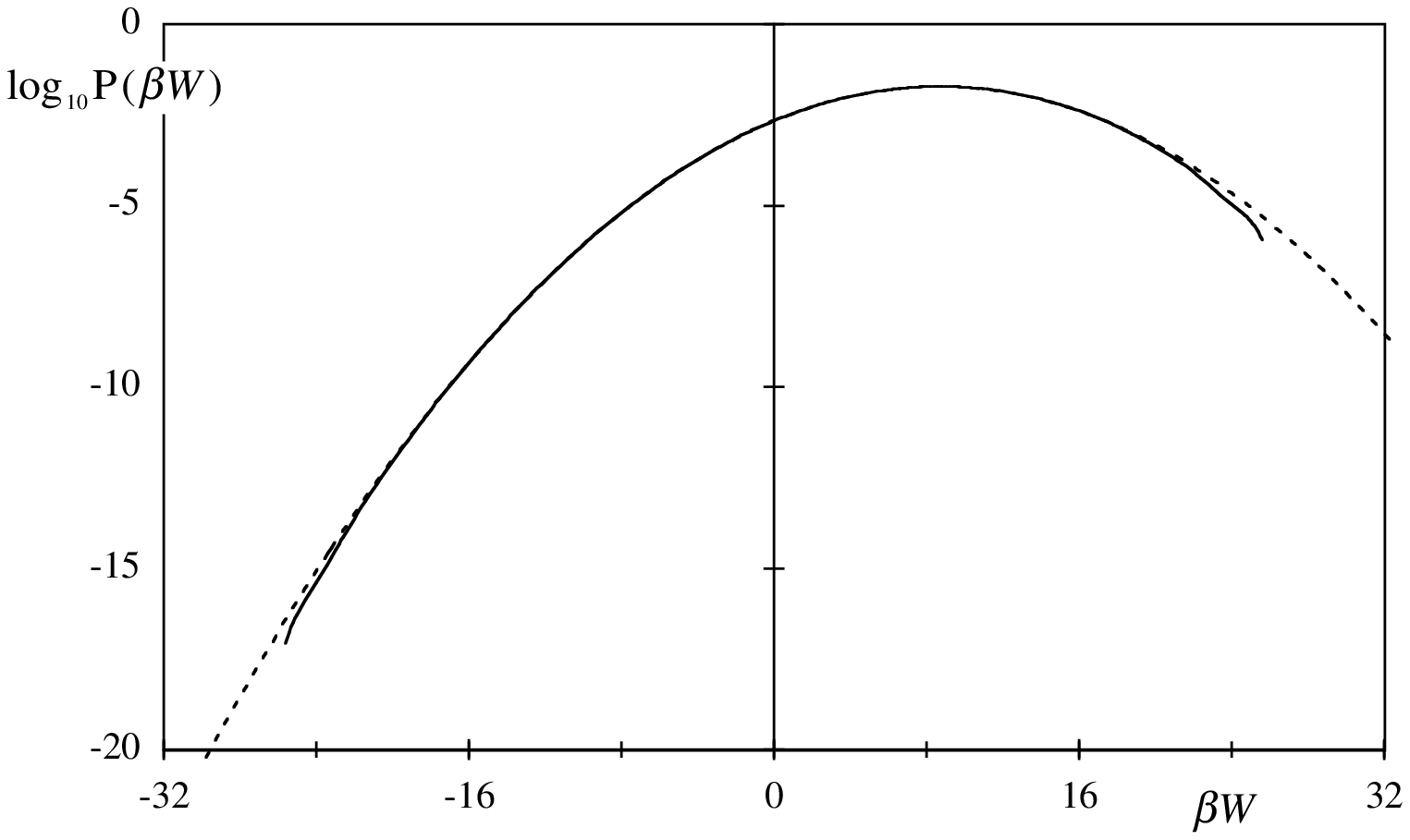, width=8cm} 
			\caption{ Work distribution (---) and Gaussian 
			approximation (- - -) for $\Delta t$ =2048, with the 
			probabilities plotted on a logarithmic scale.  The 
			Gaussian is fitted to the mean of the distribution and has 
			a variance half the mean (see Sec.~\ref{Sec:Approx}). 
			This Gaussian approximation is very accurate, even to the 
			very wings of the distribution, for times much longer than 
			the relaxation time ($\Delta t\approx 100$) of the system.  
			This same distribution is shown on a linear scale in 
			Fig.~\ref{Fig:Work}.}
		 	\label{Fig:LogWork}
	\end{figure}

	%-------- Time symmetric nonequilibrium steady states. -----------
	Systems that start and end in equilibrium are not the only ones 
	that satisfy the fluctuation theorem.  Consider again the 
	classical gas confined in a diathermal cylinder.  If the piston is 
	driven in a time symmetric periodic manner (for example, the 
	displacement of the piston is a sinusoidal function of time), the 
	system will eventually settle into a nonequilibrium steady-state 
	ensemble.  We will now require that the dynamics are entirely 
	diffusive, so that there are no momenta.  Then at any time that 
	$\lambda(t)$ is time symmetric, then the entire system is invariant 
	to a time-reversal.  We start from the appropriate nonequilibrium 
	steady-state, at a time symmetric point of $\lambda(t)$, and 
	propagate forward in time a whole number of cycles.  The 
	corresponding time reversed process is then identical to the 
	forward process, with both starting and ending in the same steady 
	state ensemble.  The entropy production for this system is odd 
	under a time reversal and the fluctuation theorem is valid.

	As a second example, consider a fluid under a constant 
	shear\cite{Evans93}.  The fluid is contained between parallel 
	walls which move relative to one another, so that $\lambda(t)$ 
	represents the displacement of the walls.  Eventually the system 
	settles into a nonequilibrium steady state.  A time reversal of 
	this steady-state ensemble will reverse all the velocities, 
	including the velocity of the walls.  The resulting ensemble is 
	related to the original one by a simple reflection, and is therefore 
	effectively invariant to the time reversal.  Again, the forward 
	process is identical to the reverse process, and the entropy 
	production is odd under a time reversal.
	
	In general, consider a system driven by a time symmetric, periodic 
	process.  [$\lambda(t)$ is a periodic, even function of time.]  We 
	require that the resulting nonequilibrium steady-state ensemble be 
	invariant under time reversal, assuming that we pick a time about 
	which $\lambda(t)$ is also symmetric.  This symmetry ensures that 
	the the forward and reverse process are essentially 
	indistinguishable, and therefore that the entropy production is 
	odd under a time reversal.  It is no longer necessary to 
	explicitly label forward and reverse processes.  ${\mathrm 
	P_F}(\omega) ={\mathrm P_R}(\omega) ={\mathrm P}(\omega)$.  For 
	these time symmetric steady-state ensembles the fluctuation 
	theorem is valid for any integer number of cycles, and can be 
	expressed as
			
		\begin{equation}
			\frac{{\mathrm P}(+\omega)}{{\mathrm P}(-\omega)} =e^{+\omega}
				\label{Eq:NSSFT} .
		\end{equation}
		
	\noindent For a system under a constant perturbation (such as 
	the sheared fluid) this relation is valid for any finite time 
	interval.
		
	\section{Long time approximations}\label{Sec:Approx}

	The steady-state fluctuation theorem, Eq.~(\ref{Eq:NSSFT}), is 
	formally exact for any integer number of cycles, but is of little 
	practical use because, unlike the equilibrium case, we have no 
	independent method for calculating the probability of a state in a 
	nonequilibrium ensemble.  The entropy production is not an easily 
	measured quantity.  However, we can make a useful approximation 
	for these nonequilibrium systems which is valid when the entropy 
	production is measured over long time intervals.

	From the relation $\langle \exp(-\omega)\rangle =1$, 
	Eq.~(\ref{Eq:One}), and the inequality $\langle \exp x\rangle 
	\geq \exp\langle x\rangle$, which follows from the convexity of 
	$e^{x}$, we can conclude that $\langle \omega\rangle \geq 0$.  On 
	average the entropy production is positive.  Because the system 
	begins and ends in the same probability distribution, the average 
	entropy production depends only on the average amount of heat 
	transferred to the bath.  $-\langle \omega\rangle =\langle \beta 
	Q\rangle \leq 0$.  On average, over each cycle, energy is 
	transferred through the system and into the heat bath  (Clausius 
	inequality).  The total heat transferred tends to increase with 
	each successive cycle.  When measurements are made over many 
	cycles, the entropy production will be dominated by this heat 
	transfer and $\omega \approx -\beta Q$.  Therefore, in the long 
	time limit the steady state fluctuation theorem, 
Eq.~(\ref{Eq:NSSFT}), can be 
	approximated as
		
		\begin{equation}
		 \lim_{\Delta t\rightarrow\infty} \frac{ {\mathrm P}(+\beta Q)} 
					{{\mathrm P}(-\beta Q)} = \exp(\beta Q) .
			\label{Eq:HeatFT}
		\end{equation}
		
	\noindent Because $-\beta Q$ is the change in entropy of the 
	bath, this heat fluctuation theorem simply ignores the 
	relatively small, and difficult to measure microscopic entropy 
	of the system. 
		
	Heat distributions for the simple computer model of a 
	nonequilibrium steady state are shown in 
	Fig.~\ref{Fig:Heat}, and the validity of the above 
	approximation is shown in Fig.~\ref{Fig:Ratio}.  As 
	expected, the heat fluctuation theorem, 
	Eq.~(\ref{Eq:HeatFT}), is accurate for times much longer 
	than the relaxation time of the system.
	
	Another approximation to the entropy production 
	probability distribution can be made in this long time 
	limit.  For long times the entropy production $\omega$ is 
	the sum of many weakly correlated values and its 
	distribution should be approximately Gaussian by the 
	central limit theorem.  If the driving process is time 
	symmetric, then ${\mathrm 
	P_F}(\omega) ={\mathrm P_R}(\omega) ={\mathrm P}(\omega)$, and the
	entropy production distribution is further constrained by 
	the fluctuation theorem itself.  The only Gaussians that 
	satisfy the fluctuation theorem have a variance twice the 
	mean, $2 \langle\omega\rangle = \bigl\langle 
	(\omega -\langle\omega\rangle)^{2}\bigr\rangle$.  This is 
	a version of the standard fluctuation-dissipation 
	relation\cite{Callen51}.  The mean 
	entropy production (dissipation) is related to the 
	fluctuations in the entropy production. If these distributions are 
	Gaussian, then the fluctuation theorem implies the Green-Kubo 
	relations for transport 
coefficients\cite{Evans93,Evans95,SearlesEvans99}. 
	However, we have not used the 
	standard assumption that the system is close to 
	equilibrium.  Instead, we have assumed that the system is 
	driven by a time symmetric process, and that the entropy 
	production is measured over a long time period.

	Gaussian approximations are shown in Figs.~\ref{Fig:Work} 
	and~\ref{Fig:LogWork} for the work distribution of the 
	simple computer model.  For times much longer than the 
	relaxation time of the system these approximations are 
	very accurate, even in the wings of the distributions.  
	This is presumably due to the symmetry imposed by the fluctuation 
	theorem.  For a nonsymmetric driving process, this symmetry 
	is broken, and the distributions will not necessarily satisfy 
	the fluctuation-dissipation relation in the long time 
	limit.  For example, see Fig.~8 of 
	Ref.~\cite{Jarzynski97b}.  Clearly these distributions 
	will be poorly approximated by the Gaussian distributions 
	considered here.

	\section{Summary}
	The  fluctuation theorem, 
	Eq.~(\ref{Eq:FT}), appears to be very general.  In this 
	paper we have derived a version that is exact for finite 
	time intervals, and which depends on the following 
	assumptions; the system is finite and classical, and 
	coupled to a set of baths, each characterized by a 
	constant intensive parameter.  The dynamics are required 
	to be stochastic, Markovian, and microscopically 
	reversible, Eq.~(\ref{Eq:MicroRev}), and the entropy 
	production, defined by Eq.~(\ref{Eq:EntropyProduction}), 
	must be odd under a time reversal.  This final condition 
	is shown to hold for two broad classes of systems, those 
	that start in equilibrium and those in a time symmetric 
	nonequilibrium steady state.  For the latter systems the 
	fluctuation theorem holds for entropy productions measured 
	over any integer number of cycles.  This generality, and 
	the existence of the nonequilibrium work relation 
	discussed in the Introduction, suggests that other 
	nontrivial consequences of the fluctuation theorem are 
	awaiting study.

	\section*{Acknowledgments} This work was initiated with 
	support from the National Science Foundation, under grant No. 
	CHE-9508336, and completed with support from the 
	U.S. Department of Energy under contract No. DE-AC03-76SF00098.
	I am indebted to C.~Jarzynski and C.~Dellago for their helpful 
	discussions, and D.~Chandler for his advice and guidance.
	
%\bibliography{fluct}

\begin{thebibliography}{10}
\bibitem{Footnote}
This definition of the entropy production is equivalent to the 
quantity
  $\Sigma$ used by Jarzynski\cite{Jarzynski97b}.
\end{thebibliography}

\end{document}